\documentclass[12pt]{article}

\usepackage{epsf}
\usepackage{graphics}
\usepackage[dvips]{graphicx}
\usepackage{epsfig}
\usepackage{amssymb}
\usepackage{caption}
\def\beq{\begin{equation}}
\def\eeq{\end{equation}}
\def\smfrac#1#2{{\textstyle {#1\over #2}}}

\def\beq{\begin{equation}}
\def\eeq{\end{equation}}
\def\ba{\begin{eqnarray}}
\def\ea{\end{eqnarray}}

\def\v8p{v_8^\prime}

\newcommand{\boldsigma}{\mbox{\boldmath$\sigma$}}
\newcommand{\boldtau}{\mbox{\boldmath$\tau$}}

\newcommand{\boldnabla}{\mbox{\boldmath$\nabla$}}

\newcommand{\ran}{\rangle}
\newcommand{\lan}{\langle}
\newcommand{\bec}{\begin{center}}
\newcommand{\enc}{\end{center}}
\newcommand{\bit}{\begin{itemize}}
\newcommand{\eit}{\end{itemize}}

\newcommand{\rta}{\rightarrow}

\newcommand{\bfA}{{\bf A}}
\newcommand{\bfB}{{\bf B}}

\newcommand{\bfO}{{\bf O}}

\newcommand{\bfS}{{\bf S}}
\newcommand{\bfT}{{\bf T}}

\newcommand{\bfq}{{\bf q}}

\newcommand{\bft}{{\bf t}}

\newcommand{\calA}{{\cal A}}

\begin{document}
\newpage
\baselineskip 16pt plus 2pt minus 2pt

\thispagestyle{empty}

\par
\topmargin=-1cm      



\vspace{0.6cm}
\begin{centering}
{\Large\bf Delta Effects in Pion-Nucleon Scattering\\
and the Strength of the Two-Pion-Exchange Three-Nucleon Interaction}

\vspace{1.0cm}

{{\bf V.R. Pandharipande}$^{1}$,
{\bf D.R.~Phillips}$^{2}$, and
{\bf U.~van Kolck}$^{3}$ }\\
\vspace{0.8cm}
{\sl $^{1}$ Department of Physics, University of Illinois,\\
Urbana, IL 61801} \\
\vspace{5.0pt}
{\sl $^{2}$Department of Physics and Astronomy, Ohio University,\\
Athens, OH 45701}\\
\vspace{5.0pt}
{\sl $^{3}$Department of Physics, University of Arizona,\\
Tucson, AZ 85721}\\
\end{centering}

\vspace{0.8cm}

\begin{center}
\begin{abstract}

\vspace*{0.1cm}

\noindent
We consider the relationship between $P$-wave $\pi N$ scattering and
the strength of the $P$-wave two-pion-exchange three-nucleon
interaction (TPE3NI). We explain why effective theories that do not
contain the delta resonance as an explicit degree of freedom tend to
overestimate the strength of the TPE3NI. The overestimation can be
remedied by higher-order terms in these ``delta-less'' theories, but
such terms are not yet included in state-of-the-art chiral EFT
calculations of the nuclear force. This suggests that these
calculations can only predict the strength of the TPE3NI to an
accuracy of $\pm 25$\%.
\end{abstract}

\vspace*{15pt}
\end{center}

\newpage

\section{Introduction}
\label{sec-intro}

A long-standing quest in hadronic physics is to relate the properties
of free pions, observed in, for instance, pion-nucleon ($\pi N$)
scattering, to those of the pions which play such a significant role
in the nuclear force.  Recently, the Nijmegen group has provided a
striking demonstration that one-pion exchange indeed provides the
longest-range component of the two-nucleon potential. They extracted,
with small error bars, the masses of the charged and neutral pions and
the couplings of pions to the nucleon from fits to the $pp$ and $np$
scattering data~\cite{KSS91}. A subsequent Nijmegen analysis
of $NN$ data then confirmed that two-pion 
exchange~\cite{deltaTPE2NP,SP76,av14,bonn,ORvK94}
gives a significant fraction of the
intermediate-range attraction in the $NN$ interaction
\cite{RTFS99}. (In some models other
mechanisms, e.g. the very broad $\sigma$ meson~\cite{PDG}, also
contribute to this attraction.)  In systems beyond $A=2$ the
three-nucleon interaction plays a subtle, but important, role.  In
this paper we focus on the Fujita-Miyazawa (FM) \cite{FM} term in the
two-pion-exchange three-nucleon interaction (TPE3NI). It appears---at
least for light nuclei---that this is the largest piece of the
three-nucleon force~\cite{vijk}.

Ideally $\pi N$ scattering data should be used to directly construct
the TPE3NI.  However, the pions that generate nuclear forces
are highly virtual. The relation between the scattering they
experience from nucleons inside the nucleus and that observed in free
space is non-trivial. To determine it, an extrapolation of the $\pi N$
amplitude from the ``physical region''---where the pion energies are 
greater than $m_\pi$---to the ``virtual region''---where pion energies are much
less than $m_\pi$---is needed. 


The delta isobar is the most prominent feature of $\pi N$ dynamics.
The delta peak in the $\pi^+p$ elastic scattering cross-section is
larger by an order of magnitude than any other~\cite{PDG}.  Therefore,
when constructing models of the $\pi N$ interaction that will be used
for the extrapolation to the virtual region it is natural to include
the delta as an explicit degree of freedom. This was the path followed
many years ago, and the leading two-pion-exchange two- and
three-nucleon potentials with an explicit delta were derived by
Sugawara \& von Hippel~\cite{deltaTPE2NP} and Fujita \&
Miyazawa~(FM)~\cite{FM}, respectively.  These two-pion-exchange $NN$
and $NNN$ potentials were recently re-derived as pieces of the more
general expressions for two- and three-nucleon forces that are
obtained when an effective field theory (EFT) with explicit delta
degrees of freedom is applied to the problem of nuclear
forces~\cite{ORvK94,vK94}. Here we discuss how the FM
potential arises in any theory with an explicit delta. Our expression
for this potential is connected to $\pi N$ scattering data through the
delta mass and the $\pi N \Delta$ coupling constant, both of which can be
determined from the $\pi N$ data.

But the highly-virtual pions exchanged in the TPE3NI have energies
much less than the delta-nucleon mass difference. This has encouraged
the development of an approach to nuclear forces that is different
from that of Sugawara \& von Hippel and Fujita \& Miyazawa. In this
approach the delta degree of freedom---along with all other $\pi N$
resonances--- is ``integrated out''. This yields an EFT in
which pions and nucleons interact in the most general
way. In this EFT $\pi N$ interactions are point-like, and are
organized as an expansion in the number of space and time
derivatives (for a review, see Ref.~\cite{BKMrev}). 
The expansion parameter is essentially
$\frac{\omega}{\Delta M}$, with $\omega$ the pion energy and
$\Delta M \equiv M_\Delta - M \approx 300$ MeV $\sim 2 m_\pi$ the
delta-nucleon mass difference. Applying this `delta-less' EFT to $\pi
N$ scattering is challenging 
(see, e.g. Ref. \cite{BKMpiN}) since the expansion parameter is, at best,
$\smfrac{1}{2}$, and the expansion breaks down completely at the delta
peak. However the expansion should converge well if $\omega \ll \Delta
M$, a condition which should have fair validity in nuclear-structure
physics. The leading contributions to $NN$ and $NNN$ potentials in
this EFT were found in Refs. \cite{OvK92} and \cite{vK94},
respectively~\footnote{The delta contributions were of course implicit
in previous dispersion-theoretical approaches \cite{DR,TM} and models
\cite{Ms,Bra},
although the correct chiral-symmetry
properties are difficult to maintain when connecting the pion-nucleon
amplitude to the potential without using field theory
\cite{friar99}.}.

We have argued that nuclear-structure physics is within the
domain of validity of both the theory with explicit deltas and the
`delta-less' EFT. We might expect then, that the two theories would give
similar results for the strength of the TPE3NI. But this turns out not
to be the case. Effective theories without an explicit delta predict a
strength for the TPE3NI that is 1.5 to 2.5 times larger than that
obtained by FM~\cite{friar99}.  Studies of the spectrum of light
nuclei with the Green's function Monte Carlo method, including 
three-nucleon interactions, favor a strength of the TPE3NI 
closer to the FM value~\cite{vijk,a9and10}~\footnote{This conclusion
is somewhat dependent on the regulator used in the three-nucleon force,
but holds definitively if one requires that the cutoffs used in the 
$NN$ and $NNN$ system be the same.}.

Here we identify the origin of this discrepancy. Parameters in the
Lagrangian of the theory with pions and nucleons alone must be
extracted from $\pi N$ scattering data. But the poor convergence of
the derivative expansion in that theory tends to contaminate
parameters extracted in this way. These parameters then appear in the
TPE3NI and result in overestimation of its strength. Within the
delta-less EFT this problem is only mitigated if many orders in the
expansion are retained.

This simple argument is presented as follows.  In
Section~\ref{sec-lag} we write down an EFT with nucleons, pions, and
explicit deltas, and compute, to leading order, both the $P$-wave $\pi
N$ scattering amplitude and the TPE3NI. In Section~\ref{sec-deltaless}
we use a theory without explicit deltas to compute the TPE3NI.  By
construction the $\pi N$ amplitudes in this theory and the theory of
Section~\ref{sec-lag} agree at $\pi N$ threshold. We show that they
differ by a factor of $\frac{4}{3}$ in their prediction for the strength of the
FM $NNN$ potential. We then discuss how this overestimation would be
remedied at higher orders in the delta-less EFT, and what the
implications of this problem are for contemporary EFT computations of
the TPE3NI.

\section{A theory with explicit deltas}
\label{sec-lag}

Although many terms contribute to $\pi N$ scattering and the
three-nucleon potential, here we focus on the delta contributions. We
do not claim that this is an accurate or complete model for either
$\pi N$ scattering or the TPE3NI, but it serves to illustrate the
point we wish to make regarding the relationship between $\pi N$ data
and the strength of the TPE3NI in delta-less EFTs. For discussions of
this relationship in the context of hadronic models, see,
e.g. Ref.~\cite{Bra}.

We consider $P$-wave $\pi N$ scattering in an effective theory with
an explicit delta degree of freedom.  We will be interested in
small pion momenta, and so we need only the leading terms in the $\pi NN$
and $\pi N \Delta$ interaction Lagrangians. 
These are: 
\ba 
{\cal L}_{\pi NN}&=& \frac{g_A}{2 f_\pi} N^\dagger \boldsigma \boldtau N
\cdot \nabla \Phi\label{eq:piNN}\\ {\cal L}_{\pi N\Delta}&=&\frac{h_A}{2 f_\pi}
(\Delta^\dagger \bfS \bfT N +\mbox{H.c.})  \cdot \nabla \Phi\, \label{eq:piND}
\ea
where $\Phi$, $N$, and $\Delta$ are the pion, nucleon and delta
fields, $f_\pi\simeq 93$ MeV is the pion decay constant, $g_A \simeq
1.29$ is the axial-vector constant that corresponds to the 
value of the (charged) $\pi NN$ coupling constant reported in
Ref.~\cite{KSS91}, $h_A \simeq 2.8$ is the
corresponding pion-nucleon-delta transition strength, and $\bfS$ and
$\bfT$ are Rarita-Schwinger transition spin and isospin operators.
Both $\bfS$ and $\bfT$ obey generalized Pauli identities of the form: 
\beq 
\bfS^{\dagger} \cdot \bfA~\bfS \cdot \bfB = 
\frac{2}{3} \bfA \cdot \bfB - \frac{1}{3} i
\boldsigma \cdot \bfA \times \bfB.
\label{gPi}
\eeq
Alternatively, one can work with the Hamiltonians
\ba
H_{\pi NN} &=& 
- \frac{f_{\pi NN}}{m_{\pi}} \boldsigma \cdot \boldnabla 
(\Phi(r) \cdot \boldtau)~, \\
H_{\pi N\Delta} &=& - \frac{f_{\pi N\Delta}}{m_{\pi}} 
\left[ \bfS \cdot \boldnabla (\Phi(r) \cdot \bfT) 
~+~ \bfS^{\dagger} \cdot \boldnabla (\Phi(r) \cdot \bfT^{\dagger}) \right]~, 
\ea
where, at this order, $f_{\pi NN}= \frac{m_\pi g_A}{2 f_\pi}$ and 
$f_{\pi N\Delta}= \frac{m_\pi h_A}{2 f_\pi}$.

\subsection{$\pi N$ scattering at low energies}

\begin{figure}[tb]
  \begin{center} 
  \mbox{\epsfig{file=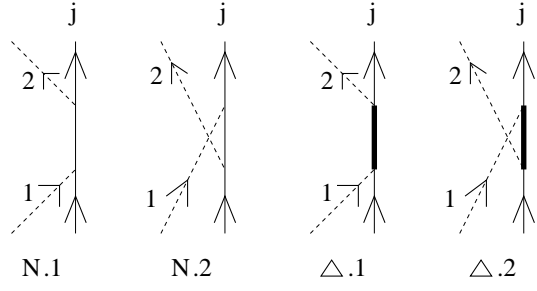,width=7truecm,angle=0}}
  \end{center}
\caption{\label{de-fig1} Four $\pi N$ scattering diagrams.
Dashed lines represent pions, solid line nucleons,
and thick solid lines delta isobars.}
\end{figure}

At leading order in small momenta these Lagrangians yield four diagrams
that contribute to $P$-wave $\pi N$ scattering. They are shown in
Fig.~\ref{de-fig1}.  Only two involve the delta. They give the
nucleon-pole-subtracted amplitude that enters the TPE3NI. Graph $\Delta.1$
is the direct---or $s$-channel---graph, and graph $\Delta.2$ is the
crossed---or $u$-channel---graph.  

We evaluate these graphs in the center-of-mass (COM) frame in which
the pion energy is $\omega$, and denote the momentum and isospin of
the initial (final) pion by $\bfq_1$ and $\bft_1$ ($\bfq_2$ and
$\bft_2$).  Since we limit ourselves to pion momenta of
the order of the pion mass the nucleon kinetic energies are smaller 
than $\omega$ by
a factor of order $m_\pi/M$, and can be neglected in this
leading-order calculation.  For the same reason we neglect the kinetic
energy of the delta.

The delta contribution to the $\pi N$ amplitude is then given by
\beq
\calA_{\pi N} = - \frac{f^2_{\pi N \Delta}}{m_{\pi}^2} 
\lan \chi'_j | \bfS_j^{\dagger} \cdot \bfq_2 \bfS_j \cdot \bfq_1 
\bfT_j^{\dagger} \cdot \bft_2 \bfT_j \cdot \bft_1 
\frac{1}{\Delta M-\omega}
+ \bfS_j^{\dagger} \cdot \bfq_1 \bfS_j \cdot \bfq_2 
\bfT_j^{\dagger} \cdot \bft_1 \bfT_j \cdot \bft_2 
\frac{1}{\Delta M+\omega} | \chi_j \ran.
\eeq
The $\chi_j$ and $\chi'_j$ are spin-isospin quantum numbers of the
nucleon before and after scattering.  

Using Eq. (\ref{gPi}), we can rewrite this amplitude as
\ba
\calA_{\pi N} &=& - \frac{f^2_{\pi N \Delta}}{m_{\pi}^2} \lan \chi'_j | 
\frac{4}{9}
\left[ \bfq_1 \cdot \bfq_2 \bft_1 \cdot \bft_2 
      -\frac{1}{4} \boldsigma_j \cdot \bfq_1 \times \bfq_2 \boldtau_j 
\cdot \bft_1 \times \bft_2 \right] 
\left( \frac{2 \Delta M}{(\Delta M)^2 - \omega^2} \right) \nonumber \\
&&\qquad 
+ i \frac{2}{9} \left[ \boldsigma_j \cdot \bfq_1 \times \bfq_2 \bft_1 
\cdot \bft_2 
+ \boldtau_j \cdot \bft_1 \times \bft_2 \bfq_1 \cdot \bfq_2 \right] 
\left( \frac{2 \omega}{(\Delta M)^2 - \omega^2} \right) | \chi_j \ran. 
\label{ApiN}
\ea

\subsection{The three-nucleon scattering amplitude}

\begin{figure}[tb]
  \begin{center} 
  \mbox{\epsfig{file=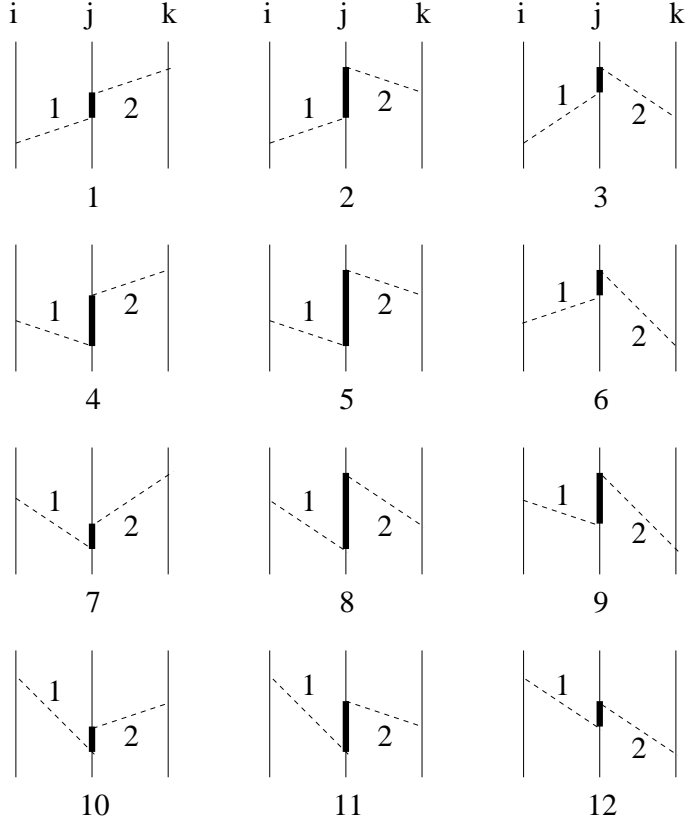,width=9truecm,angle=0}}
  \end{center}
\caption{\label{de-fig2} Twelve ``direct'' $NNN$ diagrams. Notation as
in Fig. \ref{de-fig1}.}
\end{figure}

We now turn our attention to the tree-level delta contribution in the
TPE3NI. To this end we consider the amplitude for
nucleon $i$ emitting or absorbing a pion of momentum $ \pm \bfq_1$ and
isospin $\bft_1$ and nucleon $k$ emitting or absorbing a pion of
momentum $ \pm \bfq_2$ and isospin $\bft_2$.  In ``direct'' diagrams the
pion ``1'' converts nucleon $j$ to a $\Delta$ and ``2'' reconverts it
to nucleon.  In the ``crossed'' diagrams ``2'' converts and ``1''
reconverts.  There are 12  ``direct'' and 12 ``crossed'' diagrams
in time-ordered perturbation theory.  The 12 direct diagrams are shown
in Fig.~\ref{de-fig2}.

The contribution of the direct diagrams to the three-nucleon
scattering amplitude is given by:
\ba
\calA^{direct}_{3N} &=& \frac{f^2_{\pi NN}}{m_{\pi}^2} 
           \lan \chi'_k | \boldsigma_k \cdot \bfq_2 \boldtau_k \cdot \bft_2 | 
\chi_k \ran 
           \lan \chi'_i | \boldsigma_i \cdot \bfq_1 \boldtau_i \cdot \bft_1 | 
\chi_i \ran 
           \left( \frac{1}{4 \omega_1 \omega_2} \right) 
\left[ \sum_{\alpha=1}^{12} \frac{1}{\Pi_\alpha} \right]
           \nonumber \\
&& \qquad \qquad \qquad \qquad \qquad
\times  \frac{f_{\pi N \Delta}^2} {m_{\pi}^2} \lan \chi'_j | 
  \bfS^{\dagger}_j \cdot \bfq_2 \bfS_j \cdot \bfq_1 \bfT^{\dagger}_j 
\cdot \bft_2 \bfT_j \cdot \bft_1        
| \chi_j \ran. 
\label{3namp}
\ea
Here $\chi_{i,j,k}$ and $\chi'_{i,j,k}$ denote the initial and final
spin-isospin states of nucleons $i,j$ and $k$, and $\Pi_\alpha$ is the
product of the three energy denominators in diagram $\alpha$ of
Fig.~\ref{de-fig2}.  The values of $\Pi_\alpha$ can be read off the
diagrams, 
and they are listed in
Table~\ref{table-de1}. Once again we have neglected nucleon 
and $\Delta$ kinetic energies in computing these denominators, which
is valid in our leading-order calculation.

\begin{table*}[tbp]
\begin{center}
\begin{tabular}{|c|c||c|c|}
\hline
$\alpha$   &   $-1/\Pi_\alpha$         &   $\alpha$   &  $-1/\Pi_\alpha$   \\
\hline \hline
1  & $\omega_2 \Delta M \omega_1$ &  2  & $(\Delta M + \omega_2)\Delta M\omega_1$ \\
3 & $(\Delta M + \omega_2) (\omega_1 + \omega_2) \omega_1$ & 
4 & $\omega_2 \Delta M (\Delta M + \omega_1)$ \\  
5 & $(\omega_2+\Delta M)\Delta M (\omega_1+\Delta M)$ & 
6 & $(\omega_2+\Delta M)(\omega_1+\omega_2)\omega_2$ \\
7 & $\omega_2 (\omega_1+\omega_2)(\omega_1+\Delta M)$ & 
8 & $(\omega_2 + \Delta M) (\omega_1 + \omega_2 + \Delta M) (\omega_1 + \Delta M)$ \\
9 & $(\omega_2 + \Delta M) (\omega_1 + \omega_1 + \Delta M) \omega_2$ &
10 & $\omega_1 (\omega_1 + \omega_2) (\omega_1 + \Delta M)$ \\
11 & $\omega_1 (\omega_1 + \omega_2 + \Delta M) (\omega_1 + \Delta M)$ &
12 & $\omega_1 (\omega_1 + \omega_2 + \Delta M) \omega_2$ \\
\hline
\end{tabular}
\caption{The values of $(-1/\Pi_\alpha)$ for direct diagrams. }
\label{table-de1}
\end{center}
\end{table*}

{}From Table~\ref{table-de1} we can easily verify that: 
\beq
\sum_{\alpha=1}^{12} \frac{1}{\Pi_\alpha} = \frac{-4}{\omega_1 \omega_2
\Delta M}~.  
\eeq
Substituting this in Eq.~(\ref{3namp}) gives: 
\ba
\label{3nampsum}
\calA^{direct}_{3N} &=& \frac{f^2_{\pi NN}}{m_{\pi}^2} 
           \lan \chi'_k | \boldsigma_k \cdot \bfq_2 \boldtau_k \cdot \bft_2 | 
\chi_k \ran 
           \lan \chi'_i | \boldsigma_i \cdot \bfq_1 \boldtau_i \cdot \bft_1 | 
\chi_i \ran 
           \nonumber \\
&& \times \frac{f_{\pi N \Delta}^2} {m_{\pi}^2} \lan \chi'_j | 
  \bfS^{\dagger}_j \cdot \bfq_2 \bfS_j \cdot \bfq_1 \bfT^{\dagger}_j 
\cdot \bft_2 \bfT_j \cdot \bft_1        
| \chi_j \ran \left( \frac{-1}{\omega_1^2 \omega_2^2 \Delta M} \right). 
\ea

The contribution of the crossed diagrams involves
analogous energy denominators, and 
can be calculated similarly.
The sum of direct and crossed diagrams,
\ba
\calA_{3N}  &=& \frac{f^2_{\pi NN}}{m_{\pi}^2} 
           \lan \chi'_k | \boldsigma_k \cdot \bfq_2 \boldtau_k \cdot \bft_2 | 
\chi_k \ran 
           \lan \chi'_i | \boldsigma_i \cdot \bfq_1 \boldtau_i \cdot \bft_1 | 
\chi_i \ran 
           \left( \frac{-1}{\omega_1^2 \omega_2^2 \Delta M} \right)\nonumber \\
&& \times \frac{f_{\pi N \Delta}^2} {m_{\pi}^2} \lan \chi'_j | 
  \bfS^{\dagger}_j \cdot \bfq_2 \bfS_j \cdot \bfq_1 \bfT^{\dagger}_j 
\cdot \bft_2 \bfT_j \cdot \bft_1        
~+~\bfS^{\dagger}_j \cdot \bfq_1 \bfS_j \cdot \bfq_2 \bfT^{\dagger}_j 
\cdot \bft_1 \bfT_j \cdot \bft_2  
| \chi_j \ran,
\ea
gives the Fujita-Miyazawa
potential $V^{2 \pi,FM}_{ijk}$ \cite{FM},
\ba
V^{2 \pi,FM}_{ijk} &=& \frac{f^2_{\pi NN}}{m_{\pi}^2} 
                  \left( \frac{1}{\omega_1^2 \omega_2^2} \right)
\boldsigma_k \cdot \bfq_2 \boldtau_k \cdot \bft_2 \;
\boldsigma_i \cdot \bfq_1 \boldtau_i \cdot \bft_1 \;\nonumber \\
&& \qquad \times 
\left(-\frac{f_{\pi N \Delta}^2} {m_{\pi}^2} \frac{4}{9} \frac{2}{\Delta M}
\right)
\left(\bfq_1\cdot \bfq_2 \bft_1\cdot \bft_2 
      - \frac{1}{4} \boldsigma_j \cdot \bfq_1\times \bfq_2
                   \boldtau_j \cdot \bft_1\times \bft_2 \right).
\label{eq:exact}
\ea
This result agrees with many previous re-derivations
of the FM potential, e.g. Ref.~\cite{vK94}. 
It is exact at tree level in the static limit
if the only terms in the $\pi NN$ and $\pi N \Delta$ Lagrangians
are those in Eqs.~(\ref{eq:piNN}) and (\ref{eq:piND}).

\section{Relation to theories without explicit deltas}
\label{sec-deltaless}

We now attempt to find a more direct connection between $\pi N$
scattering data and $V^{2 \pi,FM}_{ijk}$---one that does not invoke
the delta as an explicit degree of freedom. Such attempts have been
reviewed in Ref.~\cite{friar99} whose notation we follow below.

A key aspect of this connection is that $\pi N$ scattering involves
pions with $\omega \sim m_\pi$, while in $V^{2 \pi,FM}_{ijk}$ we have
$\omega \sim m_\pi^2/M$. (The typical nucleon momentum in the nucleus
is of order the pion mass, and the pion energy is then smaller by a
factor $m_\pi/M$.) Since we have already been neglecting terms
suppressed by $m_\pi/M$ we take $q_1^0=q_2^0=0$. Given this kinematics, 
the three-nucleon potential can be written:
\beq
\bar{V}^{2 \pi}_{ijk} = \frac{f_{\pi NN}^2}{m_{\pi}^2} \frac{\boldsigma_i \cdot
\bfq_1~\boldsigma_k \cdot \bfq_2}{\omega_1^2 \omega_2^2}
\left[-F_j^{\alpha \beta} \tau_i^{\alpha} \tau_k^{\beta} \right]~,
\eeq
where $\omega_i\equiv \sqrt{\bfq_i^2 +m_\pi^2}$ comes from the
pion propagators and 
\beq 
-F_j^{\alpha \beta} = \delta^{\alpha \beta}
\left[ a + b~\bfq_1 \cdot \bfq_2 + c(q_1^2 + q_2^2) \right] -
d(\tau_j^{\gamma} \epsilon^{\alpha \beta \gamma} \boldsigma_j \cdot
\bfq_1 \times \bfq_2)
\label{eq:deltalessform}
\eeq 
is the Born-subtracted $\pi N$ subamplitude.  The first term is due to
$S$-wave scattering, the second gives the anticommutator part of the
TPE3NI, the third is zero, and the fourth gives the commutator part.
The first term is very small in the context of $V^{2 \pi}_{ijk}$
\cite{vijk}, and it is zero in the present model.

The crucial point, then, is the determination of the coefficients $b$
and $d$.  In a theory without explicit delta fields, they are fitted
to $\pi N$ data near threshold.  If we lived in a world where there
were no contributions to $\pi N$ scattering other than from the $s$-
and $u$-channel delta and nucleon poles, comparing
Eq.~(\ref{eq:deltalessform}) and Eq.~(\ref{ApiN}) shows that a fit to
threshold $\pi N$ data would result in
\ba 
b = 4 d = - \frac{f^2_{\pi N
\Delta}}{m_{\pi}^2}~\frac{4}{9} \left( \frac{2 \Delta M}{(\Delta M)^2 -
m_{\pi}^2} \right)~. \label{b}
\ea
The TPE3NI corresponding to this amplitude is given by:
\beq
\bar{V}^{2 \pi}_{ijk} = \frac{f^2_{\pi NN}}{m_{\pi}^2}
~\frac{1}{\omega_1^2 \omega_2^2} 
~\boldsigma_i \cdot \bfq_1~ \boldsigma_k \cdot \bfq_2
~\boldtau_i \cdot \bft_1~ \boldtau_k \cdot \bft_2~\bfO_j^{\pi N}~. 
\label{eq:Vbar}
\eeq
The factor $1/\omega_1^2
\omega_2^2$ comes from the pion propagators, and the factors
besides $\bfO_{j}^{\pi N}$
describe the coupling of the pions to the nucleons $i$ and $k$.
The $\pi N$ interaction is described by:
\beq
\bfO_j^{\pi N} = b
\left(\bfq_1 \cdot \bfq_2~\bft_1 \cdot \bft_2 
            - \frac{1}{4}~\boldsigma_j \cdot \bfq_1 \times \bfq_2~\boldtau_j 
\cdot \bft_1 \times \bft_2\right),
\label{eq:O}
\eeq 
with $b$ given by Eq.~(\ref{b}).  Of course, this is just the usual FM
form, but with specific choices for the coefficients $b$ and
$d$.  

\subsection{The problem}

Comparing the $\bar{V}_{ijk}^{2 \pi}$ in Eq. (\ref{eq:Vbar}) 
with the the ``exact''
result for our model ($V_{ijk}^{2 \pi,FM}$ of Eq.~(\ref{eq:exact})) we find
that they are the same apart from the crucial fact that the strength
of the interaction in the ``delta-less'' theory has the factor
$\frac{2 \Delta M}{(\Delta M)^2 - m_{\pi}^2}$, instead of the
$\frac{2}{\Delta M}$ of the ``exact'' result.  Since $\Delta M
\simeq 2 m_{\pi}$, these factors are $\simeq \frac{4}{3m_\pi}$ and
$\simeq \frac{1}{m_\pi}$, respectively. One way to understand this
result is to realize that the direct term for the $\pi N$ scattering
amplitude in Eq.~(\ref{ApiN}) and Fig.~\ref{de-fig1} is evaluated at
the energy of a real pion, and so has the energy denominator $\Delta
M-m_{\pi}$ for low-momentum pions. This denominator is half of the
average denominator, $\Delta M$, of the diagrams in Fig.~\ref{de-fig2}
that contribute to the TPE3NI.  The crossed pion term mitigates this
discrepancy, but not enough to cure the problem. Ultimately, 
{\it the $\bar{V}_{ijk}^{2 \pi}$ that is extracted ``directly'' from $\pi N$
scattering data is too strong by a factor of 4/3.}

The difference between $\bar{V}^{2 \pi}_{ijk}$ and $V^{2 \pi,FM}_{ijk}$ 
is of order $\left(\frac{m_\pi}{\Delta M}\right)^2$.  It
will vanish in the limit $\Delta M \gg m_{\pi}$, which
includes the chiral limit $m_{\pi} \rta 0$.  However, in the context
of the nuclear many-body problem $m_{\pi}$ {\bf is not small}.  The range
of OPEP is comparable to the mean inter-nucleon spacing in nuclei, and
the energies required to excite nucleons to isobar states such as the
delta are not much larger than $m_{\pi}$.

Of course, in the real world there are contributions to the $\pi N$
amplitude other than the two graphs we have considered here.  Also $b$
and $d$ will probably be determined from data that are not exactly at
threshold.  While we cannot say {\it a priori} in which direction
these effects go, fitting $\pi N$ data at higher energies will
presumably only make the extrapolation problem worse.

Parts of this problem have been understood for a long time, but, 
as discussed in
the introduction, the prevailing folklore has been that an EFT without
explicit deltas could still work well in nuclei, because the relevant
energies in nuclear-structure physics are much smaller than
$\Delta M$.  However, the poor convergence of the EFT without explicit
deltas for $\pi N$ scattering
affects the TPE3NI because $b$ and $d$ are not calculated from 
first principles; instead they are fitted to threshold $\pi N$ data.  This
necessitates an extrapolation from pion energies $\omega \sim m_\pi$
to the energies of the highly-virtual pions in the TPE3NI, which are of
order $\frac{m_\pi^2}{M}$.  
This extrapolation takes place over an energy range that is
sizable compared to the radius of convergence of the ``delta-less''
theory---$\Delta M$.

Here we have explicitly considered the implications of such an
extrapolation for the three-nucleon potential, but other few-nucleon
potentials (including the two-nucleon force) will be afflicted by the
same problem. All use $\pi N$ parameters that are potentially
contaminated in a similar way.  Such contamination will occur in all
EFTs for low-energy hadronic physics which contain only pion and
nucleon degrees of freedom.

\subsection{The solution}

In a theory with explicit deltas this extrapolation is under much
better control, since the pion-energy dependence of the $\pi N$
amplitude is better reproduced. In contrast, at leading order in the
``delta-less'' theory the coefficients of the two operators in 
${\bf O}_j^{\pi N}$ are energy independent, and so the value
extracted for them at threshold, where $\omega=m_\pi$, is used 
in the TPE3NI, where $\omega \simeq 0$.

But at higher orders in this EFT additional corrections to the $\pi N$
amplitude, and in particular to the two operators in ${\bf O}_j^{\pi
N}$, enter.  To see what form this higher-order energy dependence
would take, we expand the result (\ref{ApiN}) in powers of
$\left(\frac{\omega}{\Delta M}\right)^2$. The first correction to the
leading-order results for $b$ and $d$ (\ref{b}) occurs at
$O[\left(\frac{\omega}{\Delta M}\right)^2]$. The form of ${\bf
O}_j^{\pi N}$ is now:
\begin{equation}
\bfO_j^{\pi N} = 
\left(b + \tilde{b} \omega^2\right)
\left(\bfq_1 \cdot \bfq_2~\bft_1 \cdot \bft_2 -
\frac{1}{4}~\boldsigma_j \cdot \bfq_1 \times \bfq_2~\boldtau_j \cdot
\bft_1 \times \bft_2\right). \label{eq:btilde}
\end{equation}
In the EFT, terms such as $\tilde{b} \omega^2$ and 
$\tilde{d} \omega^2$ appear in the Lagrangian as pion-nucleon
interactions with time derivatives. 
We must fit $\pi N$ data over a range of pion energies to
determine both $b$  and $\tilde{b}$.
If, once again, we imagine living in a world where the true answer was
given by Eq.~(\ref{ApiN}), then fitting the form
(\ref{eq:btilde}) to reproduce (\ref{ApiN}) in the region
around $\omega=m_\pi$ yields:
\begin{eqnarray}
b&=&-\frac{4}{9}\frac{f_{\pi N \Delta}^2}{m_\pi^2}
\left(\frac{2 \Delta M}{(\Delta M)^2 - m_\pi^2}\right)\left(1 - \frac{m_\pi^2}{(\Delta M)^2 - m_\pi^2}\right);\\
\tilde{b}&=&-\frac{4}{9}\frac{f_{\pi N \Delta}^2}{m_\pi^2}
\frac{2 \Delta M}{((\Delta M)^2 - m_\pi^2)^2}.
\label{eq:new}
\end{eqnarray}
Note that at $\pi N$ threshold this gives exactly the same result for
$\bfO_j^{\pi N}$ as in Eq.~(\ref{eq:O}). However, extrapolating to
$\omega=0$ now yields a TPE3NI that has an additional factor of 
$(1 - \frac{m_\pi^2}{(\Delta M)^2 - m_\pi^2})$ in its strength.  
If we set $\Delta M=2 m_\pi$,
this gives an overall factor of $\frac{8}{9m_\pi}$, instead of the 
$\frac{1}{m_\pi}$
found in the ``exact'' calculation with explicit deltas. This
means that in the theory {\it without} explicit deltas the ``exact''
factor $\frac{1}{m_{\pi}}$ is being built up as:
\begin{equation}
\frac{1}{m_\pi}=\left(1 - \frac{1}{3} + \frac{1}{9} + \ldots\right)
\frac{4}{3 m_\pi},
\end{equation}
a series that converges moderately quickly. 

To summarize: in the theory without explicit deltas it is important to
realize that the factor $\frac{4}{3 m_\pi}$ obtained by fitting
$\pi N$ ``data'' with the leading-order form (\ref{eq:O}) is not the
final answer.  This result will change when higher-order terms are
incorporated in the theory and used to improve the extrapolation from
$\omega \simeq m_\pi$ to $\omega \simeq 0$. We can estimate the size of
such terms based on our knowledge that the convergence will be
governed by the parameter $\frac{m_\pi}{\Delta M}$, and that---due to
crossing---only even terms in this expansion can appear in $b$ and
$d$. The leading-order result should therefore be quoted as:
\begin{equation}
b=-\frac{4}{9} \frac{f_{\pi N \Delta}^2}{m_\pi^2} 
\frac{4}{3 m_\pi} \left[1 \pm \left(\frac{m_\pi}{\Delta M}\right)^2 \right].
\end{equation}
More conservative error bars are certainly acceptable, but the
$\approx 25$\% we have chosen is the minimum permissible
theoretical error that can be assigned to $b$ when it is extracted in
the theory without explicit deltas. Such an error bar turns out to be
consistent with the ``exact'' answer for $b$ in the simple model
considered here.

\section{Conclusion}

We have shown that theories without an explicit delta tend to
overestimate the delta contribution to the TPE3NI.  This is because
there is an error in the leading-order computation of the
three-nucleon potential in the ``delta-less'' theory.  The error is
$\sim$ 25\%, and it is necessary to include terms suppressed by
$\left(\frac{\omega}{\Delta M}\right)^2$ in the EFT to reduce it.
The inclusion of other higher-order effects, such as nucleon recoil
and dispersive effects for intermediate-state deltas, may make the
extrapolation error smaller than we found, but it seems
unlikely that it will completely remove the difficulty.

Unfortunately this problem is present in the state-of-the-art N$^3$LO
chiral EFT computation of $NN$ and $NNN$
potentials~\cite{moregermans}. The terms that ameliorate the
overestimation appear in ${\cal L}_{\pi N}^{(4)}$, and so will not
enter the chiral EFT nuclear force until N$^4$LO. Computing the two-
and three-nucleon potentials to this (or higher) order will take
considerable effort. It may well be that an EFT with explicit deltas
is simply a more efficient tool than one without.  In fact, the first
studies in nuclear EFT \cite{ORvK94,vK94} included diagrams with
intermediate deltas in their calculation of the nuclear force.  The
drawback of such a treatment is that in order to fix parameters one
must analyze data around the delta resonance, which necessitates a
resummation of the delta self-energy.  Only recently has a power
counting been devised that allows a systematic EFT treatment of
effects in this kinematic region~\cite{dansdelta}.

The delta-less EFT has also found difficulties with certain $\pi N$
parameters that are large because the effects of the integrated-out
delta are encoded there. In Ref.~\cite{germans} Epelbaum {\it et al.}
argued that there is a cancellation of delta-excitation and $\pi
\rho$-exchange contributions in nuclear forces. This motivated their
use of $NN$ and $NNN$ potentials containing $\pi N$-interaction
parameters smaller than those extracted from chiral analyses of $\pi
N$ scattering data. We stress that the reduction in strength of the
$NNN$ force we have discussed here is not based on such an
argument. It is independent of details of nuclear dynamics at the
distance scale $1/m_\rho$.

So, until the theory with explicit delta degrees of freedom is further
developed, or delta-less theories can be extended to higher order,
the $\pi N$ parameters used in the $NNN$ potential should be
viewed as only loosely constrained by $\pi N$ data. Furthermore, EFT
extractions of $\pi N$ parameters from $NN$ data (see,
e.g. Ref.~\cite{RTFS99}) and from $\pi N$ data (see, e.g.
Ref.~\cite{BKMpiN}) can be expected to give results that differ by
amounts of order $\left(\frac{m_\pi}{\Delta M}\right)^2$.

\bigskip

\section*{Acknowledgments}
We thank Evgeny Epelbaum and Steve Pieper for useful discussions.  UvK
and DRP thank the INT program ``Microscopic Nuclear Structure Theory''
for hospitality during the inception of this work.  UvK also thanks
the Nuclear Theory Group at the University of Washington for
hospitality while part of this work was carried out.  This research
was supported in part by the US NSF under grant PHY 03-55014 (VRP), by
the US DOE under grants DE-FG02-93ER40756 and DE-FG02-02ER41218 (DRP), and
DE-FG02-04ER41338 (UvK), and by an Alfred P. Sloan Fellowship (UvK).


\begin{thebibliography}{99}

\bibitem{KSS91} 
R. A. M. Klomp, V. G. J. Stoks and J. J. de Swart, 
Phys. Rev. {\bf C44}, R1258 (1991).

\bibitem{deltaTPE2NP}
H. Sugawara and F. von Hippel, 
Phys. Rev. {\bf 172}, 1764 (1968).

\bibitem{SP76}
R. A. Smith and V. R. Pandharipande, 
Nucl. Phys. {\bf A256}, 327 (1976).

\bibitem{av14}
R. B. Wiringa, R. A. Smith, and T. L. Ainsworth, 
Phys. Rev. {\bf C29}, 1207 (1984). 

\bibitem{bonn}
R. Machleidt, Adv. Nucl. Phys. {\bf 19}, 189 (1989). 

\bibitem{ORvK94}
C. Ord\'{o}\~{n}ez, L. Ray, and U. van Kolck,
Phys. Rev. Lett. {\bf 72}, 1982 (1994);
Phys. Rev. {\bf C53}, 2086 (1996).

\bibitem{RTFS99} 
M. C. M. Rentmeester, R. G. E. Timmermans, J. L. Friar, and J. J. de Swart, 
Phys. Rev. Lett. {\bf 82}, 4992 (1999); 
M.~C.~M.~Rentmeester, R.~G.~E.~Timmermans, and J.~J.~de Swart,
Phys.\ Rev.\ {\bf C67}, 044001 (2003).

\bibitem{PDG}
Particle Data Group, Phys. Rev. {\bf D66}, 266 (2002).  

\bibitem{FM}
J.-I. Fujita and H. Miyazawa, 
Prog. Theor. Phys. {\bf 17}, 360 (1957). 

\bibitem{vijk} 
S. C. Pieper, V. R. Pandharipande, R. B. Wiringa, and J. Carlson, 
Phys. Rev. {\bf C64}, 014001 (2001). 

\bibitem{vK94}
U. van Kolck, 
Phys. Rev. {\bf C49}, 2932 (1994).

\bibitem{BKMrev}
V.~Bernard, N.~Kaiser, and U.-G.~Meissner,
Int.\ J.\ Mod.\ Phys.\ {\bf E4}, 193 (1995).

\bibitem{BKMpiN}
V.~Bernard, N.~Kaiser, and U.-G.~Meissner,
Nucl.\ Phys.\ {\bf A615}, 483 (1997);
N. Fettes and U.-G.~Meissner,
Nucl.\ Phys.\ {\bf A693}, 693 (2001).

\bibitem{OvK92}
C. Ord\'{o}\~{n}ez and U. van Kolck, 
Phys. Lett. {\bf B291}, 459 (1992);
J. L. Friar,
Phys. Rev. {\bf C60}, 034002 (1999).

\bibitem{DR}
W. N. Cottingham, M. Lacombe, B. Loiseau, J. M. Richard, and R. Vinh Mau, 
Phys. Rev. {\bf D8}, 800 (1973);
A. D. Jackson, D. O. Riska, and B. Verwest,
Nucl.\ Phys.\ {\bf A249}, 397 (1975).

\bibitem{TM}
S. A. Coon, M. D. Scadron, P. C. McNamee, B. R. Barrett, D. W. E. Blatt, 
and B. H. J. McKellar, 
Nucl. Phys. {\bf A317}, 242 (1979).

\bibitem{Ms}
L. S. Celenza, A. Pantziris, and C. M. Shakin, 
Phys. Rev. {\bf C46}, 2213 (1992);
C. A. da Rocha and M. R. Robilotta,
Phys. Rev. {\bf C49}, 1818 (1994); 
Phys. Rev. {\bf C52}, 531 (1995);
Nucl. Phys. {\bf A615}, 391 (1997);
J.-L. Ballot, M. R. Robilotta, and C. A. da Rocha,
Int. J. Mod. Phys. {\bf E6}, 83 (1997);
Phys. Rev. {\bf C57}, 1574 (1998).

\bibitem{Bra}
H. T. Coelho, T. K. Das, and M. R. Robilotta, 
Phys. Rev. {\bf C28}, 1812 (1983); 
M. R. Robilotta and H. T. Coelho,
Nucl. Phys. {\bf A460}, 645 (1986);
T.~Y.~Saito and I.~R.~Afnan,
Few Body Syst.\  {\bf 18}, 101 (1995);
T.~Y.~Saito and J.~Haidenbauer,
Eur.\ Phys.\ J.\  {\bf A7}, 559 (2000);
D.~P.~Murphy and S.~A.~Coon, Few Body Syst.\  {\bf 18}, 73 (1995).

\bibitem{friar99}
J. L. Friar, D. H\"uber, and U. van Kolck,
Phys. Rev. {\bf C59}, 53 (1999).

\bibitem{a9and10} 
S. C. Pieper, K. Varga, and R. B. Wiringa, 
Phys. Rev. {\bf C66}, 044310 (2002). 

\bibitem{moregermans}
D.~R.~Entem and R.~Machleidt,
Phys.\ Rev.\  {\bf C68}, 041001 (2003);
E. Epelbaum, W. Gl\"ockle, and U.-G. Mei{\ss}ner, 
[arXiv:nucl-th/0405048].

\bibitem{dansdelta}
V. Pascalutsa and D.~R. Phillips, 
Phys.\ Rev.\  {\bf C67}, 055202 (2003);
V.~Pascalutsa and D.~R.~Phillips,
Phys.\ Rev.\  {\bf C68}, 055205 (2003).

\bibitem{germans}
E. Epelbaum, A. Nogga, W. Gl\"ockle, H. Kamada, U.-G. Mei{\ss}ner, 
and H. Wita\l a,
Eur. Phys. J. {\bf A15}, 543 (2002).

\end{thebibliography}
\end{document}